\documentstyle[amsmath,amssymb,epsfig,12pt]{article}

\begin{document}
\setlength{\parskip}{0.45cm}
\setlength{\baselineskip}{0.75cm}
%
%
%
\begin{titlepage}
\setlength{\parskip}{0.25cm}
\setlength{\baselineskip}{0.25cm}
\begin{flushright}
DO-TH 2005/03\\
\vspace{0.2cm}
March 2005
\end{flushright}
\vspace{1.0cm}
\begin{center}
\Large
{\bf Radiatively Generated Isospin Violations}
\\\Large{\bf in the Nucleon and the NuTeV Anomaly}
\vspace{1.5cm}

\large
M.\ Gl\"uck, P.\ Jimenez--Delgado, E.\ Reya\\
\vspace{1.0cm}

\normalsize
{\it Universit\"{a}t Dortmund, Institut f\"{u}r Physik,}\\
{\it D-44221 Dortmund, Germany} \\
\vspace{0.5cm}

\vspace{1.5cm}
\end{center}

\begin{abstract}
\noindent Predictions of isospin asymmetries of valence and sea distributions 
are presented which are generated by QED leading ${\cal{O}}(\alpha)$ 
photon bremsstrahlung effects.  Together with isospin violations arising
from nonperturbative hadronic sources (such as quark and target mass
differences) as well as with even a conservative contribution from a
strangeness asymmetry ($s\neq \bar{s}$), the discrepancy between the
large NuTeV `anomaly' result for $\sin^2\theta_W$ and the world average
of other measurements is removed. 
\end{abstract}
\end{titlepage}


The NuTeV collaboration recently reported \cite{ref1} a measurement of
the Weinberg angle $s_W^2\equiv \sin^2\theta_W$ which is approximately
three standard deviations above the world average \cite{ref2} of other
electroweak measurements.  Possible sources for this discrepancy 
(see, for example, \cite{ref3,ref4,ref5,ref6,ref7}) include, among other
things, isospin-symmetry violating contributions of the parton 
distributions in the nucleon, i.e., nonvanishing $\delta q_v$ and
$\delta\bar{q}$ defined via
\begin{eqnarray}
\delta u_v(x,Q^2) & = & u_v^p(x,Q^2)-d_v^n(x,Q^2)\nonumber\\
\delta d_v(x,Q^2) & = & d_v^p(x,Q^2)-u_v^n(x,Q^2)
\end{eqnarray}
where $q_v=q-\bar{q}$ and with analogous definitions for $\delta\bar{u}$
and $\delta\bar{d}$.  The valence asymmetries $\delta u_v$ and 
$\delta d_v$ were estimated within the nonperturbative framework of the
bag model \cite{ref4,ref5,ref8,ref9,ref10} and resulted in a reduction
of the above mentioned discrepancy by about 30\%.  It should be 
emphasized that these nonperturbative charge symmetry violating
contributions arise predominantly through mass differences $\delta m =
m_d-m_u$ of the struck quark and from target mass corrections
related to $\delta M=M_n-M_p$.

The additional contribution to the valence isospin asymmetries stemming
from radiative QED effects was presented recently \cite{ref11}.
Following the spirit of this publication we shall evaluate $\delta q_v$
and $\delta\bar{q}$ in a slightly modified way based on the approach
presented in \cite{ref12,ref13} utilizing the QED ${\cal{O}}(\alpha)$
evolution equations for $\delta q_v(x,Q^2)$ and $\delta\bar{q}(x,Q^2)$
induced by the photon radiation off the (anti)quarks.
To {\em{leading}} order in $\alpha$  we have
\begin{eqnarray}
\frac{d}{d\ln Q^2}\, \delta u_v(x,Q^2) & = & \frac{\alpha}{2\pi}\int_x^1
 \frac{dy}{y}\,P\left(\frac{x}{y}\right) u_v(y,Q^2)\nonumber\\
\frac{d}{d\ln Q^2}\, \delta d_v(x,Q^2) & = & -\frac{\alpha}{2\pi}\int_x^1
 \frac{dy}{y}\,P\left(\frac{x}{y}\right) d_v(y,Q^2)
\end{eqnarray}
with $P(z) = (e_u^2-e_d^2)P_{qq}^{\gamma}(z) = (e_u^2-e_d^2)
\left(\frac{1+z^2}{1-z}\right)_+$,
and similar evolution equations hold for the isospin asymmetries of 
sea quarks $\delta\bar{u}(x,Q^2)$ and $\delta\bar{d}(x,Q^2)$.  Notice
that the addition \cite{ref11,ref14} of further terms proportional to
$(\alpha/2\pi)e_q^2 P_{q\gamma}*\gamma$ to the r.h.s. of (2) would
actually amount to a subleading ${\cal{O}}(\alpha^2)$ contribution since
the photon distribution $\gamma(x,Q^2)$ of the nucleon is of 
${\cal{O}}(\alpha)$ \cite{ref15,ref16,ref17,ref18,ref19,ref20}. 
We integrate (2) as follows:
\begin{eqnarray}
\delta u_v(x,Q^2) & = & \frac{\alpha}{2\pi} 
      \int_{m_q^2}^{Q^2}d\ln q^2 
       \int_x^1\frac{dy}{y}\,\, 
         P\left(\frac{x}{y}\right) u_v(y,\, q^2)\nonumber\\
\delta d_v(x,Q^2) & = & -\frac{\alpha}{2\pi}
      \int_{m_q^2}^{Q^2}d\ln q^2 
       \int_x^1\frac{dy}{y}\,\, 
         P\left(\frac{x}{y}\right) d_v(y,\, q^2)
\end{eqnarray}
and similarly for $\delta\bar{u}$ and $\delta\bar{d}$ utilizing the usual
isospin symmetric leading--order (LO) parton distributions $q_v(x,q^2)$
and $\bar{q}(x,q^2)$ of the dynamical (radiative) parton model
\cite{ref21}. The current quark mass $m_q$ being the usual kinematical
lower bound for a photon emitted by a quark -- similar to the electron
mass $m_e$ for a photon radiated off an electron \cite{ref22}.  Here 
we conservatively choose $m_q=10$ MeV, i.e., of the order of the current
quark masses \cite{ref2}.  The parton distributions at 
$q^2<\mu_{\rm LO}^2$ in (3), where $\mu_{\rm LO}^2=0.26$ GeV$^2$ is
the input scale in \cite{ref21}, are taken to equal their values at the 
perturbative input scale $\mu_{\rm LO}^2$, 
$\stackrel{(-)}{q}\!\!(y,\, q^2\leq \mu_{\rm LO}^2) =
\stackrel{(-)}{q}\!\!(y,\, \mu_{\rm LO}^2)$, i.e.\ are `frozen'.

The resulting valence isospin asymmetries $\delta u_v$ and $\delta d_v$
at $Q^2=10$ GeV$^2$ are presented in Fig.\ 1 where they are compared
with the corresponding nonperturbative bag model results \cite{ref5},
with the latter ones being of entirely different origin, i.e., arising
dominantly through the mass differences $\delta m$ and $\delta M$.
As can be seen, our radiative QED predictions and the bag model estimates
are comparable for $\delta u_v$ but differ considerably for $\delta d_v$.
It should furthermore be noted that, although our method differs 
somewhat from that in \cite{ref11}, our resulting $\delta q_v(x,Q^2)$
turn out to be quite similar, as already anticipated in \cite{ref11}.

Going beyond the results in \cite{ref4,ref5,ref8,ref9,ref10} and
\cite{ref11} we present in Fig.\ 2 our estimates for the isospin
violating sea distributions for $\delta\bar{u}$ and $\delta\bar{d}$
at $Q^2 = 10$ GeV$^2$.  Similar results are obtained for the LO CTEQ4
parton distributions \cite{ref23} where also valence--like sea 
distributions are employed at the input scale $Q_0^2=0.49$ GeV$^2$,
i.e., $x\bar{q}(x,Q_0^2)\to 0$ as $x\to 0$.  Such predictions may be
tested by dedicated precision measurements of Drell--Yan and DIS
processes employing neutron (deuteron) targets as well.

Turning now to the impact of our $\delta\!\! \stackrel{(-)}{q}\!\!\!(x,Q^2)$
on the NuTeV anomaly, we present in Table I the implied corrections 
$\Delta s_W^2$ to $s_W^2$ evaluated according to
\begin{equation}
\Delta s_W^2 =  \int_0^1 F
  [s_W^2,\,\delta\!\!\stackrel{(-)}{q};\,x]\, 
    \delta\!\!\stackrel{(-)}{q}\!\!\!(x,Q^2)\, dx
\end{equation}
at $Q^2\simeq 10$ GeV$^2$, appropriate for the NuTeV experiment. The
functionals $F[s_W^2,\, \delta\!\!\stackrel{(-)}{q};\, x]$
are presented in \cite{ref3} according to the experimental methods
\cite{ref1} used for the extraction of $s_W^2$ from measurements of
\begin{equation}
R^{\nu(\bar{\nu})}(x,Q^2)\equiv d^2\sigma_{NC}^{\nu(\bar{\nu})N}
  (x,Q^2)/d^2\sigma_{CC}^{\nu(\bar{\nu})N}(x,Q^2)\, .
\end{equation}
Since the isospin violation generated by the QED ${\cal{O}}(\alpha)$
correction is such as to remove more momentum from up--quarks than
down--quarks, as is evident from Fig.~1, it works in the right 
direction to reduce the NuTeV anomaly \cite{ref1}, i.e., 
$\sin^2\theta_W = 0.2277\pm 0.0013\pm 0.0009$ as compared to the world
average of other measurements \cite{ref2} $\sin^2\theta_W = 0.2228(4)$.
Also shown in Table I are the {\em{additional}} contributions
to $\Delta s_W^2$ stemming from the nonperturbative hadronic bag
model calculations \cite{ref4,ref5,ref8,ref9,ref10} where isospin
symmetry violations arise predominantly through the quark and 
target mass differences $\delta m$ and $\delta M$, respectively, as
mentioned earlier.  These contributions are comparable in size to our
radiative QED results.

Although the NuTeV group \cite{ref1} has taken into account several
uncertainties in their original analysis due to a nonisoscalar target,
higher--twists, charm production, etc., they have disregarded, besides
isospin violations, effects caused by the strange sea asymmetry
$s\neq \bar{s}$.  
\begin{table}
\begin{center}
\renewcommand{\arraystretch}{1.5}
\begin{tabular}{l||ccccc}
$\Delta s_W^2$ & $\delta u_v$ & $\delta d_v$ & $\delta\bar{u}$ 
    & $\delta\bar{d}$ & total\\
\hline
QED & -0.00071 & -0.00033 & -0.000019 & -0.000023 & -0.0011\\
\hline
bag & -0.00065 & -0.00081 & --- & --- & -0.0015\\
\hline
\end{tabular}
\end{center}
\caption
{The QED corrections to $\Delta s_W^2$ evaluated 
according to (4) using (3).  The nonperturbative bag model estimates
\cite{ref9} are taken from \cite{ref5}; different nonperturbative
approaches give similar results \cite{ref5}.}
\end{table}
Recent nonperturbative estimates 
\cite{ref7,ref24,ref25,ref26} resulted in sizeable contributions to
$\Delta s_W^2$ similar to the ones in Table I.  As a conservative
estimate we use \cite{ref25} $\Delta s_W^2|_{\rm strange} = -0.0017$.
With the results in Table I, the {\em{total}} correction 
therefore becomes
\begin{eqnarray}
\Delta s_W^2|_{\rm total} & = & \Delta s_W^2|_{\rm QED} + 
     \Delta s_W^2|_{\rm bag} + \Delta s_W^2|_{\rm strange}\nonumber\\
& = & -0.0011\,\,\, - \,\,\,0.0015 \,\,\,-\,\,\,0.0017\nonumber\\
& = & -0.0043\, .
\end{eqnarray}
Thus the NuTeV measurement (`anomaly') of 
$\sin^2\theta_W = 0.2277(16)$ will be shifted to $\sin^2\theta_W =
0.02234(16)$ which is in agreement with the standard value 0.2228(4).

Finally, it should be mentioned that, for reasons of simplicity,
it has become common (e.g.\ \cite{ref6,ref7,ref11,ref24,ref26}) to use
the Paschos--Wolfenstein relation \cite{ref27} for an isoscalar target,
$R_{\rm PW}^-=\frac{1}{2}-s_W^2$, for estimating the corrections
discussed above,
\begin{equation}
R^-\equiv\frac{\sigma_{\rm NC}^{\nu N}-\sigma_{\rm NC}^{\bar{\nu}N}}
              {\sigma_{\rm CC}^{\nu N}-\sigma_{\rm CC}^{\bar{\nu}N}}
  = R_{\rm PW}^- + \delta R_I^- +\delta R_s^-\, ,
\end{equation}
instead of the experimentally directly measured and analyzed ratios
$R^{\nu(\bar{\nu})}$ in (5), where \cite{ref3}
\begin{equation}
\delta R_I^-\simeq \left(\frac{1}{2}-\frac{7}{6}s_W^2\right)
 \frac{\delta U_v-\delta D_v}{U_v+D_v}\,\,,\quad
\delta R_s^-\simeq -\left( 1-\frac{7}{3}s_W^2\right)\frac{S^-}{U_v+D_v}
\end{equation}

\noindent 
with $Q_v(Q^2) = \int_0^1 x\,q_v(x,Q^2)\, dx,\quad\quad \delta\, Q_v(Q^2) = 
\int_0^1 x\,\delta\, q_v(x,Q^2)\, dx$ and $S^-(Q^2) = \int_0^1x[s(x,Q^2) 
-\bar{s}(x,Q^2)]\,  dx$.
(Note that the correct expressions for {\em{both}} 
$\delta R_I^-$ and $\delta R_s^-$ have been presented only in 
\cite{ref3}).  Our radiative QED results in Fig.\ 1 imply 
$\delta U_v = -0.002226$ and $\delta D_v = 0.000890$ which, together 
with $U_v+D_v = 0.3648$, give 
\mbox{$\Delta s_W^2|_{\rm QED}=\delta R_I^-|_{\rm QED}= -0.002$}
according to (8), whereas the correct value in Table I is only 
{\em{half}} as large.  Similar overestimates are obtained
for the nonperturbative (hadronic) bag model results \cite{ref5}.
Furthermore, the frequently used \cite{ref6,ref7,ref24,ref26} 
expression for $\delta R_s^-$ in (8) due to a strangeness asymmetry
represents already a priori an overestimate since it results from
treating naively the CC transition 
$\stackrel{(-)}{s}\, \rightarrow\,
\stackrel{(-)}{c}$
without a kinematic suppression factor for massive charm production
\cite{ref3}.  Nevertheless one obtains
$\Delta s_W^2|_{\rm strange} = \delta R_s^-=-0.0021$ using
\cite{ref25} $S^-=0.00165$, instead of $\Delta s_W^2|_{\rm strange}=
-0.0017$ in (6), as derived from (4). Therefore the $\delta R_{I,s}^-$
in (8) should be avoided, in particular $\delta R_I^-$, and the shift
in $s_W^2$ should rather be evaluated according to (4) corresponding
to the actual NuTeV measurements \cite{ref1}. 

To summarize, we evaluated the modifications 
$\delta\!\!\!\stackrel{(-)}{q}\!\!\!(x,Q^2)$ to the standard isospin 
symmetric
parton distributions due to QED ${\cal{O}}(\alpha)$ photon bremsstrahlung
corrections.  Predictions are obtained for the isospin violating
valence $\delta q_v$ and sea $\delta\bar{q}$ distributions ($q = u,d$)
within the framework of the dynamical (radiative) parton model.
For illustration we compared our radiative QED results for the
isospin asymmetries $\delta u_v(x,Q^2)$ and $\delta d_v(x,Q^2)$ with
nonperturbative bag model calculations where the violation of isospin
symmetry arises from entirely {\em{different}} (hadronic)
sources, predominantly through quark and target mass differences.
Taken together, these two isospin violating effects reduce already
significantly the large NuTeV result for $\sin^2\theta_W$.  Since,
besides isospin asymmetries, the NuTeV group has also disregarded
possible effects caused by a strangeness asymmetry ($s\neq\bar{s}$)
in their original analysis \cite{ref1}, we have included a recent
conservative estimate of the $s\neq\bar{s}$ contribution to 
$\Delta\sin^2\theta_W$ as well.  Together with the isospin violating
contributions (cf.(6)), the discrepancy between the large result for
$\sin^2\theta_W$ as derived from deep inelastic $\nu(\bar{\nu})N$ data
(NuTeV `anomaly') and the world average of other measurements is
entirely removed.

\newpage
\begin{figure}
\epsfig{file=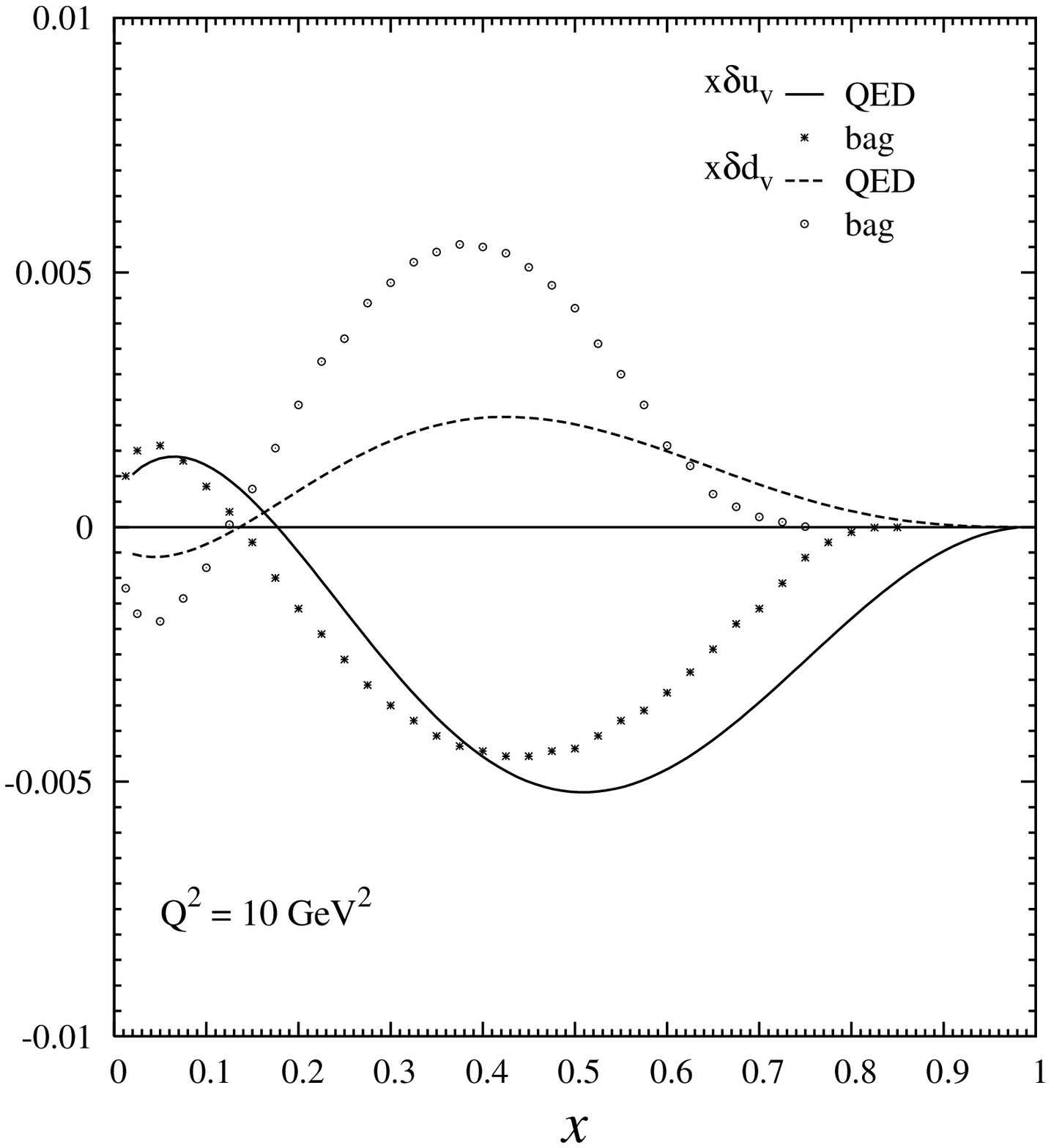,width=\textwidth}
\caption{The isospin violating `majority' $\delta u_v$
      and  `minority' $\delta d_v$ valence quark distributions at
      $Q^2=10$ GeV$^2$ as defined in (1).  Our radiative QED 
      predictions are calculated according to (3).  The bag model 
      estimates are taken from Ref.~\cite{ref5}.}
\end{figure}

\newpage
\begin{figure}
\epsfig{file=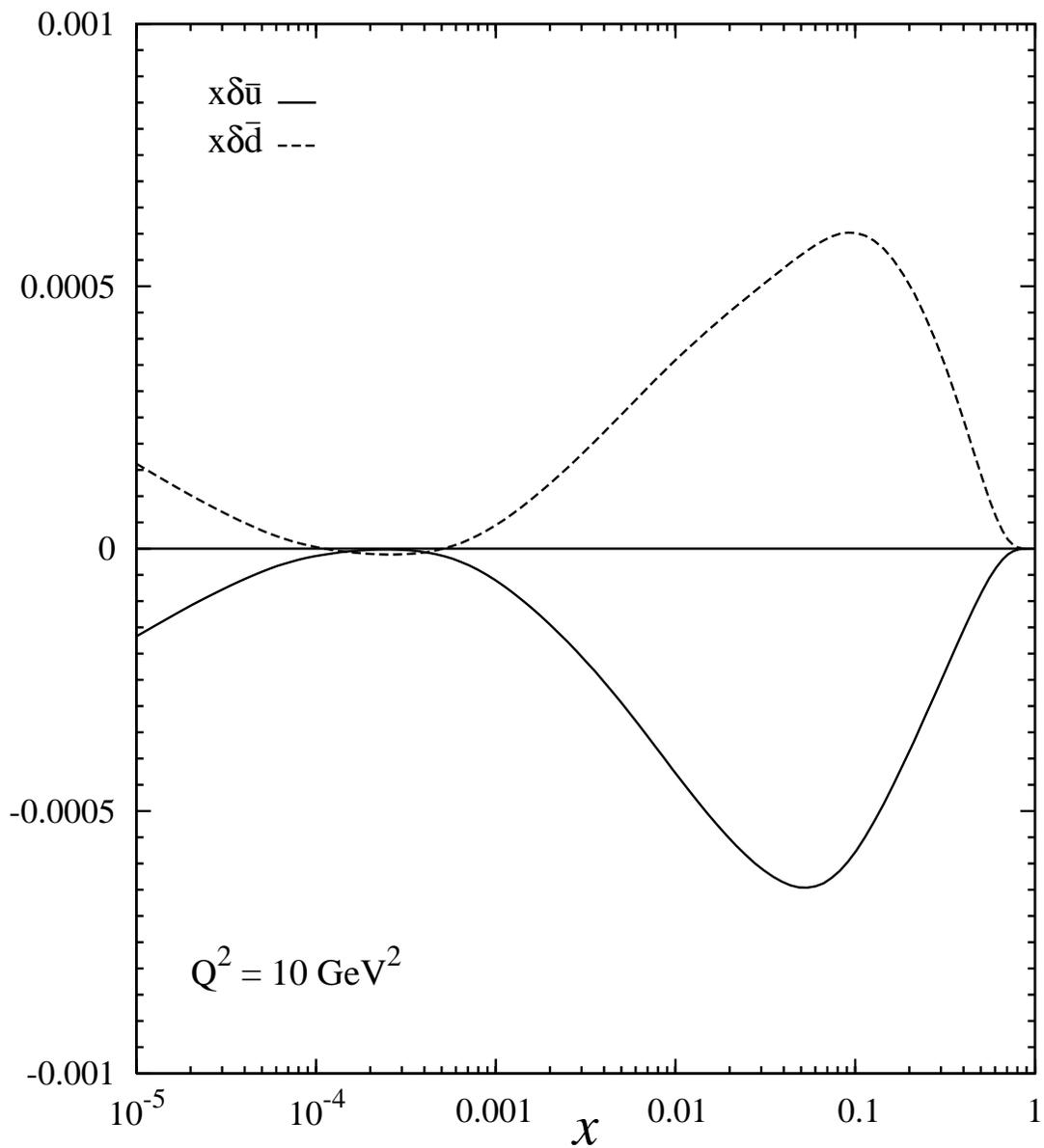,width=\textwidth}
\caption{The isospin violating sea distributions 
      $\delta\bar{u}$ and $\delta\bar{d}$ at $Q^2=10$ GeV$^2$ as
      defined in (1) with $u_v$, $d_v$ replaced by $\bar{u},\, 
      \bar{d}$.  The QED predictions are calculated
      according to (3) with $u_v,\, d_v$ replaced by 
      $\bar{u},\, \bar{d}$.}
\end{figure}


\end{document}